\documentstyle[12pt]{article}
\begin{document}

\setlength{\textheight}{1.2\textheight}
\begin{titlepage}
\nopagebreak
\begin{center}
{\large \bf Space-time localisation with quantum fields}\\
\end{center}
\vfill
\begin{center}
{\bf Marc-Thierry Jaekel$^a$ and Serge Reynaud$^b$} \\
\end{center}

\begin{flushleft}
$(a)$Laboratoire de Physique Th\'{e}orique
 de l'ENS\footnote{Unit\'e
propre du Centre National de la Recherche Scientifique, \\
associ\'ee \`a l'Ecole Normale Sup\'erieure et \`a l'Universit\'e
de Paris Sud.}(CNRS-UPS),\\
24 rue Lhomond, F75231 Paris Cedex 05\\
$(b)$Laboratoire Kastler Brossel\footnote{Unit\'e
de l'Ecole Normale Sup\'erieure et de l'Universit\'e Pierre et Marie Curie,\\
 associ\'ee au Centre National de la Recherche
Scientifique.}(UPMC-ENS-CNRS), case 74,\\
4 place Jussieu, F75252 Paris Cedex 05\\
\end{flushleft}
\vfill

\begin{abstract}
We introduce observables associated with the space-time position of a
quantum point defined by the intersection of two light pulses. The time
observable is canonically conjugated to the energy. Conformal symmetry of
massless quantum fields is used first to build the definition of these
observables and then to describe their relativistic properties under frame
transformations. The transformations to accelerated frames of the space-time
observables depart from the laws of classical relativity. The Einstein laws
for the shifts of clock rates and frequencies are recovered in the quantum
description, and their formulation provides a conformal metric factor
behaving as a quantum observable.
\end{abstract}
\begin{flushleft}
\vfill
LPTENS 96/30
\qquad {\it to appear in} Physics Letters A

\end{flushleft}

\end{titlepage}

The formulation of relativistic theories relies on the concept of events
localised in space-time. As a seminal example, the relativistic property of
time was introduced by Einstein when he questionned the notion of
simultaneity and defined synchronisation through the transfer of light
pulses between remote clocks \cite{Einstein05}. A key point in this
operational approach is the possibility of associating space-time positions
with events such that emission or reception of a pulse. Such positions
necessarily are physical observables which have to be distinguished from
coordinate parameters on a map of space-time. In particular, the
relativistic properties of these observables result from a symmetry of the
laws of physics, namely their Lorentz invariance. A naive identification of
these properties with map transformations could only be an oversimplifying
and misleading shortcut. This is the reason why the interpretation of
relativity, as embodied in symmetry principles, has been repeatedly upheld
against the more common and substantially different interpretation of
relativity, as rooted in covariance principles \cite{Norton93}.

While the previous arguments refer to classical theories of relativity, the
concept of space-time localisation is known to raise challenging issues in
the context of quantum theory \cite{NWSW}. In standard quantum formalism,
time is never treated as an operator, and thus has a quite different
description from that of space position. Moreover, the very notion of
quantum fluctuations of time remains a matter of debate, since the formalism
does not provide a precisely stated energy-time commutation relation which
would assert that the fourth Heisenberg inequality effectively constrains
time and energy fluctuations \cite{Jammer74}. Differences in the description
of space and time variables and the subsequent inconsistency between the
formalism of quantum theory and the geometric description of space-time are
also known to be knotty points in attempts to include gravity in quantum
theory \cite{Rovelli91}.

The basic idea underlying the present work is that space-time observables
certainly belong to the quantum domain, like clocks used for time definition
and electromagnetic signals used for synchronisation. Metrological
considerations support this idea, since the definition of space-time units
is now rooted in atomic physics \cite{TimeFrequency}. These arguments imply
that relativity theory has to be consistent with a quantum description of
space-time observables. They also point to the need for a clear distinction
between two commonly confused notions of time. On one hand, time is the
basic evolution parameter used to write dynamical laws and conservation
laws. On the other hand, time is the physical variable delivered by a clock
when a given physical event occurs. The former notion of time is clearly
distinct from any spatial variable while the latter one is an extension of
the concept of localisation in space to that of localisation in time. It is
this latter definition of time, and not the former one, which is mixed with
that of space by relativistic transformations. As already noticed, these
relativistic transformations have to be considered as laws of physics and
not simply as direct consequences of map transformations. In particular, the
shifts of space-time observables under transformations to accelerated frames
\cite{Einstein07} cannot be deduced from covariance properties associated
with the corresponding map transformations.

In this respect, the case of uniform acceleration requires a specific
attention, since there exist conformal coordinate transformations which fit
accelerated motion and still preserve propagation equations of
electromagnetic field \cite{BC09}. This property is a particular case of
conformal invariance of massless field theories \cite{FRW62}. The
propagation of such fields is not sensitive to a conformal variation of the
metric tensor, that is a change of space-time scale preserving the velocity
of light \cite{MG80}. Since this conformal symmetry enlarges the symmetry
built on Lorentz invariance and inertial motions, it may be expected that it
allows to determine the relativistic properties of space-time observables
for uniformly accelerated observers, in the same manner as Lorentz
invariance for inertial observers. Indeed, preliminary results in this
direction have been obtained for the problem of clock synchronisation
performed through the transfer of a light pulse. The time reference encoded
in the pulse, which has to be shared by two remote observers, must be a
quantity preserved by field propagation. Such a quantity may actually be
built from conformal generators associated with the field state and its
relativistic properties may then be deduced from the conformal algebra \cite
{PRL96}.

Following this constructive approach, we show in the present letter that
quantum space-time observables may be associated with a physical event
defined by the intersection of two light pulses and that these definitions
fulfill correct quantum and relativistic properties. First, these
observables will be found to obey canonical conjugation relations with
momentum operators. In particular, time will be defined as a quantum
observable with fluctuations obeying an energy-time Heisenberg relation.
Moreover, the definitions given in this letter will describe time and space
observables in an explicitly Lorentz invariant manner. Then, the
relativistic properties of space-time and energy-momentum observables as
seen by inertial or accelerated observers will be derived from the conformal
algebra. These properties, which will generalise covariance rules to the
quantum domain, will be seen to depart from the statements inherited from
classical relativity.

Before coming to a fully quantum mechanical discussion, we will introduce
the necessary distinction between space-time observables and coordinate
parameters in a classical context. This will allow us to show how space-time
observables are constructed from the generators associated with conformal
symmetry and how their relativistic properties under frame transformations
are obtained from the same symmetry. We first define Lie transformations,
which represent changes of frame as deformations of the coordinate map
\begin{equation}
x^\mu \stackrel{a}{\rightarrow }\overline{x}^\mu =x^\mu +\varepsilon
_a\delta _a^\mu (x)  \label{map}
\end{equation}
where $\varepsilon _a$ is an infinitesimal number and $\delta _a^\mu $ a
polynomial function of coordinate parameters $x$. The commutator between two
frame transformations is thus represented by the difference between the
images of a point $x$ through the composed deformations ($a\circ b$) and ($%
b\circ a$), which is evaluated as $\varepsilon _a\varepsilon _b\delta
_{(a,b)}^\mu (x)$ where $\delta _{(a,b)}^\mu $ is the Lie commutator
\begin{equation}
\delta _{(a,b)}^\mu =\delta _b^\nu \partial _\nu \delta _a^\mu -\delta
_a^\nu \partial _\nu \delta _b^\mu   \label{comm}
\end{equation}
We shall assume that frame transformations are performed around an inertial
frame and raise or lower tensor indices by using the Minkowski tensor $\eta
_{\mu \nu }$. However, these transformations give in general rise to a
change of the metric tensor
\begin{equation}
g_{\mu \nu }=\eta _{\mu \nu }\stackrel{a}{\rightarrow }\overline{g}_{\mu \nu
}=\eta _{\mu \nu }-\varepsilon _a\left( \partial _\mu \delta _{a\nu
}+\partial _\nu \delta _{a\mu }\right)   \label{metric}
\end{equation}

For Lorentz transformations, the metric tensor is unchanged. For conformal
transformations, its change reduces to a point-dependent rescaling
\begin{eqnarray}
g_{\mu \nu }=\eta _{\mu \nu }\stackrel{a}{\rightarrow }\overline{g}_{\mu \nu
} &=&\eta _{\mu \nu }\left( 1+2\varepsilon _a\lambda _a(x)\right)   \nonumber
\\
2\lambda _a\eta _{\mu \nu } &=&-\partial _\mu \delta _{a\nu }-\partial _\nu
\delta _{a\mu }  \label{cfac}
\end{eqnarray}
Conformal transformations are thus associated with the preservation of the
velocity of light and, consequently, of the propagation of massless fields
\cite{BC09}. The commutator (\ref{comm}) of conformal transformations is
still a conformal transformation, and the set of these commutators
constitutes the conformal algebra which characterises the symmetry
properties of the field theory \cite{Itzykson}. The associated conserved
quantities, that is the generators of the symmetries, are used in the
following to characterise the field states while the conformal algebra
determines the transformations of these quantities and therefore the
transformations of the field states. The prime role is thus played by the
conformal generators which correspond to translations ($P_\nu $), rotations (%
$J_{\nu \rho }$), dilatation ($D$) and conformal transformations to
uniformly accelerated frames ($C_\nu $) and are given respectively by the
following deformations \cite{FRW62}
\begin{eqnarray}
\delta _{P_\nu }^\mu (x)=\eta _\nu ^\mu  &\qquad &\lambda _{P_\nu }(x)=0
\nonumber \\
\delta _{J_{\nu \rho }}^\mu (x)=\eta _\nu ^\mu x_\rho -\eta _\rho ^\mu x_\nu
&\qquad &\lambda _{J_{\nu \rho }}(x)=0  \nonumber \\
\delta _D^\mu (x)=x^\mu  &\qquad &\lambda _D(x)=-1  \nonumber \\
\delta _{C_\nu }^\mu (x)=2x_\nu x^\mu -\eta _\nu ^\mu x_\rho x^\rho  &\qquad
&\lambda _{C_\nu }(x)=-2x_\nu   \label{cct}
\end{eqnarray}
$\eta _\nu ^\mu $ denotes a Kronecker symbol.

We then consider a classical light ray defined as a dispersionless field
distribution running along a light ray parametrised as
\begin{equation}
x^\mu =\xi ^\mu +p^\mu \sigma \qquad p^\mu p_\mu =0  \label{traj}
\end{equation}
where $\xi ^\mu $ represents any origin on the ray, $p^\mu $ a light-like
momentum vector and $\sigma $ an affine parameter along the ray. This
parametrisation relies on the assumption of a dispersionless distribution
both in momentum and position spaces. This assumption, clearly inconsistent
with the principles of quantum physics, will be released in the following.
In this restricted context however, it allows us to write in a simple manner
the conformal generators
\begin{equation}
\Delta =p_\nu \delta ^\nu (x)  \label{gen}
\end{equation}
The conservation laws for these generators then appear as a direct
consequence of definitions (\ref{cfac}) and (\ref{traj})
\begin{equation}
\frac{{\rm d}\Delta }{{\rm d}\sigma }=p_\nu p^\mu \ \partial _\mu \delta
^\nu =-\lambda \ p_\mu p^\mu =0  \label{conserv}
\end{equation}
We can now make clear that the evolution parameter $\sigma $ is distinct
from any conceivable notion of time observable. As a matter of fact, the
values of the generators $\Delta $ are conserved, i.e. independent of $%
\sigma $, on each light ray whereas they clearly vary with variables
describing the position of the ray in space-time. In particular, they are
changed when the origin $\xi _\mu $ of the light ray is displaced
\begin{equation}
{\rm d}\Delta =p_\nu \ {\rm d}\xi ^\mu \ \partial _\mu \delta ^\nu (\xi
)\neq 0  \label{transl}
\end{equation}
except when the displacement ${\rm d}\xi ^\mu $ is parallel to the momentum
and thus preserves the ray. The last equation gives the change ${\rm d}%
\Delta $ of conformal generators when the light ray is translated in a given
reference frame, but also when the frame is translated, provided that the
signs are carefully taken care of.

We now describe the ray transformations in a more general manner. We first
characterize the various rays by the values of the generators (\ref{gen}).
This characterisation is intrinsic since it does not rely upon an arbitrary
parametrisation such as the one of equation (\ref{traj}). Conversely, such a
parametrisation may be derived, if necessary, from the values of the
momentum and rotation generators. The set of classical light rays is then
mapped into itself by the conformal transformations, and this mapping is
described by the conformal algebra. If one denotes $\Delta _a$ the generator
associated with a transformation and $\Delta _b$ a conserved quantity
associated with a ray, the change of $\Delta _b$ under the transformation $%
\Delta _a$ may be read as $\varepsilon _a\Delta _{(a,b)}$ where $\Delta
_{(a,b)}$ is the conformal generator given by the commutator (\ref{comm})
\begin{equation}
\Delta _{(a,b)}=p_\nu \delta _{(a,b)}^\nu =\delta _b^\mu \partial _\mu
\Delta _a-\delta _a^\mu \partial _\mu \Delta _b  \label{ConfAlg}
\end{equation}
The particular case (\ref{transl}) of changes ${\rm d}\Delta $ under
translations is now embodied in the commutator between $P_\mu $ and $\Delta $%
. Since a commutator is antisymmetric in the exchange of its two arguments,
this change is also the opposite of the change of momentum $P_\mu $ under
the action of $\Delta $.

The discussion restricted up to now to the case of a classical light ray
shows that the conformal symmetry allows to characterise the rays as well as
to describe the effects of frame transformations. We may stress that a field
state containing a single light ray is associated with a geometrical line
rather than with a point. In contrast, a field state comprising two rays can
be used to obtain the position of a localised event defined as the point of
intersection of the two lines. We therefore consider now a state built with
two field pulses coinciding at a space-time position $X^\mu $ and
propagating in different directions defined by momenta $p_{\pm }^\mu $. Each
of the two rays may be parametrised by equation (\ref{traj}) with both
origins chosen at the coincidence point $X^\mu $. Conserved quantities (\ref
{gen}), defined as sums of the contributions of the two rays, thus read
\begin{equation}
\Delta =P_\nu \delta ^\nu (X)  \label{genn}
\end{equation}
where $P_\nu $ is the total energy-momentum of the field state $\sum_{\pm
}p_{\nu \pm }$.

Since the elementary rays have different propagation directions, the mass
associated with the field state
\begin{equation}
M^2=P_\nu P^\nu   \label{defM2}
\end{equation}
no longer vanishes. The expressions of the rotation and dilatation
generators $J^{\mu \nu }$ and $D$ may therefore be inverted to obtain the
space-time position of the coincidence point
\begin{eqnarray}
J^{\mu \nu } &=&P^\mu X^\nu -P^\nu X^\mu \qquad D=P_\mu X^\mu   \nonumber \\
X^\mu  &=&\frac{P^\mu }{M^2}D-\frac{P_\nu }{M^2}J^{\mu \nu }  \label{defX}
\end{eqnarray}
We may emphasize that the connection between the definition of an observable
space-time position $X^\mu $ and the massive character of the state has a
simple physical interpretation, in the spirit of the discussion about
localisation already presented. A vanishing mass $M$ corresponds to a field
state with a single propagation direction which cannot be associated with a
localised event. In contrast, a non vanishing mass reveals that the state
contains different light rays which intersect and thus define a point in
space-time. It is also worth stressing that the observable position $X$ is
built from conserved quantities and, consequently, does not evolve due to
field propagation
\begin{equation}
\frac{{\rm d}X^\mu }{{\rm d}\sigma }=0
\end{equation}

This confirms that space-time observables conceptually differ from
coordinate parameters, for example from those associated with field pulses
running along trajectories. In particular, the temporal component $X^0$ must
not be confused with the affine parameter $\sigma $. As already discussed, a
variation of $\sigma $ describes pulse propagation along the trajectories
while a variation of $X^0$ corresponds to a translation of the trajectories
and therefore of the time observable associated with the coincidence event.
More generally, changes of frame are described by the conformal algebra (\ref
{ConfAlg}). Notice that the commutators between the conformal generators may
also be written as Poisson brackets which generalise the Lie commutators
initially defined for map deformations to observables \cite{deWitt}
\begin{equation}
\Delta _{(a,b)}\equiv \left( \Delta _a,\Delta _b\right) =\frac{\partial
\Delta _a}{\partial X^\mu }\frac{\partial \Delta _b}{\partial P_\mu }-\frac{%
\partial \Delta _a}{\partial P_\mu }\frac{\partial \Delta _b}{\partial X^\mu
}  \label{poisson}
\end{equation}

We come now to the quantum mechanical discussion where the distinction
between observables and evolution parameter will become even more striking
than in the classical context. Space-time observables will indeed appear as
quantum operators, whereas $\sigma $ will remain a classical evolution
parameter. In the quantum discussion, the field states exhibit momentum and
position dispersions as a consequence of Heisenberg relations. Conformal
generators are defined as integrals of operators $T_{\mu \nu }$ representing
components of the stress tensor
\begin{equation}
\Delta =\int_\sigma {\rm d}\Sigma \ T_{\mu 0}(x)\ \delta ^\mu (x)
\label{Gen}
\end{equation}
The symbol $\int_\sigma {\rm d}\Sigma $ denotes an integral over a
space-like surface at constant coordinate parameter $\sigma $. The
energy-momentum densities $T_{\mu 0}$ are normally ordered so that they
vanish in vacuum. Notice that the conformal transformations (\ref{cct}) do
not only preserve the propagation equation, but also the definition of
vacuum \cite{QSO95} and of particle number \cite{BJP95}. Noether's theorem
asserts \cite{Itzykson} that the generators (\ref{Gen}) are preserved by
field propagation like the classical ones (see (\ref{conserv})). The
canonical field commutators and the definition of stress tensor are such
that the quantum commutators of the conformal generators (\ref{Gen})
identify with the Lie commutators \cite{deWitt}
\begin{equation}
\frac 1{i\hbar }\left[ \Delta _a,\Delta _b\right] =\left( \Delta _a,\Delta
_b\right)   \label{Comm}
\end{equation}
For this reason, we will further use the notation $\left( \ ,\ \right) $
rather than $\frac 1{i\hbar }\left[ \ ,\ \right] $ for writing the
commutation relations.

Now, observable positions in space-time may be defined from the conformal
generators. As a matter of fact, centers of inertia of the energy-momentum
distribution may be obtained by inverting expressions of rotation and
dilatation generators $J^{\mu \nu }$ and $D$ as in classical equations (\ref
{defX})
\begin{eqnarray}
J^{\mu \nu } &=&P^\mu \cdot X^\nu -P^\nu \cdot X^\mu \qquad D=P_\mu \cdot
X^\mu  \nonumber \\
X^\mu &=&\frac{P^\mu }{M^2}\cdot D-\frac{P_\nu }{M^2}\cdot J^{\mu \nu }
\label{DefX}
\end{eqnarray}
We have taken care of non commutativity of quantum observables by
introducing a symmetrised product represented by the $"~\cdot ~"$ symbol. We
have also assumed that the mass (\ref{defM2}) associated with the field
state does not vanish. The definition (\ref{DefX}) of space-time positions
associated with the field state is quite analogous to Einstein's classical
definition of spatial positions \cite{Einstein06}. However, it involves not
only the rotation generators $J^{\mu \nu }$ but also the dilatation
generator $D$. As a result, an observable time is defined together with
space positions. Furthermore, the definition (\ref{DefX}) holds in the
quantum domain, with the particularly important outcome that the space-time
observables are canonically conjugated to energy-momentum operators
\begin{equation}
\left( P^\mu ,X^\nu \right) =-\eta ^{\mu \nu }  \label{canonical}
\end{equation}
Hence, the commutators of functions of energy-momentum and space-time
observables may equivalently be written as Poisson brackets (\ref{poisson}).
The space-time observables are not themselves conformal generators and thus
do not belong to the conformal algebra. However, the canonical commutation
relations (\ref{canonical}) involve the enveloping algebra, that is the
structure built on polynomial functions of the conformal generators. In this
sense, canonical commutation relations can be considered as embodied in
conformal symmetry. It is worth emphasizing that an energy-time Heisenberg
relation is now obtained besides momentum-space relations of standard
quantum formalism. Furthermore, these relations enter a Lorentz invariant
description.

Relations (\ref{DefX}) imply that the generators $J_{\mu \nu }$ and $D$ have
a classical form in terms of the observable $X$ defined as center of the
energy-momentum distribution. This is related to the fact that the
corresponding deformations $\delta ^\mu $ are linear functions of $x$. Since
transformations to accelerated frames correspond to quadratic functions, the
generators $C_\nu $ will differ from the corresponding classical form
because of dispersions associated with Heisenberg relations \cite{PRL96}. It
is thus natural to write the various generators as
\begin{eqnarray}
\Delta  &=&P_\mu \cdot \delta ^\mu (X)+\widehat{\Delta }  \nonumber \\
\widehat{P}_\nu  &=&\widehat{J}_{\nu \rho }=\widehat{D}=0\qquad \widehat{C}%
_\nu \neq 0  \label{Gencorr}
\end{eqnarray}
The first contribution to $\Delta $ has a classical form to be compared with
(\ref{genn}). The second contribution $\widehat{\Delta }$ is thus defined as
a correction to the classical expression which differs from $0$ only for
transformations to accelerated frames. We may then check by inspection of
the conformal algebra (\ref{ConfAlg}) that the commutators of any generator $%
\Delta $ with a translation generator $P_\mu $ belong to the set of
generators which have a classical form
\begin{equation}
\left( \Delta ,P_\mu \right) =P_\nu \cdot \partial _\mu \delta ^\nu (X)
\label{DP}
\end{equation}
This proves that the corrections $\widehat{\Delta }$ always commute with the
momentum operators
\begin{equation}
\left( \widehat{\Delta },P_\mu \right) =\frac{\partial \widehat{\Delta }}{%
\partial X^\mu }=0
\end{equation}
Therefore the non vanishing corrections $\widehat{C}_\nu $ may be written in
terms of the momentum operators and of the Casimir invariants of the
conformal algebra. Dimensional analysis implies that $\widehat{C}_\nu $
scales as the inverse of a momentum and is thus a non-linear expression of
momentum. To specify the argument, we will assume that the field state
consists of two intersecting light rays with identical dispersions. In this
case, the correction $\widehat{C}_\nu $ has the simple form
\begin{equation}
\widehat{C}_\nu =\alpha \frac{P_\nu }{M^2}  \label{Casimir}
\end{equation}
where $\alpha $ is a Casimir invariant of the conformal algebra. It is
related to the Casimir invariants associated with each light ray \cite{PRL96}
and is positive with a minimal magnitude of the order of $\hbar ^2$.

Having defined space-time observables in terms of the conformal generators,
we come now to the next stage of the analysis, devoted to their
transformations under changes of frame. The commutator $\left( \Delta ,P_\mu
\right) $ characterises the momentum change under the frame transformation
associated with $\Delta $, as it results from a comparison with classical
relations (\ref{transl},\ref{poisson}). Equations (\ref{DP}) thus mean that
this momentum change has precisely the form expected for the shift of a
vector field in classical differential geometry. This is true not only for
Lorentz transformations, but also for dilatations and transformations to
accelerated frames. In particular, the Einstein prediction of a position
dependent momentum change for transformations to accelerated frames \cite
{Einstein07} is recovered in the context of quantum theory. Moreover, the
dependence of this change in terms of observable positions is the same as
for the classical expression written in terms of coordinate parameters.

This perfect matching between quantum and classical laws still holds for the
shift of space-time observables $X^\mu $ under translations, rotations and
dilatations, but not under transformations to accelerated frames. Using
equation (\ref{Gencorr}), we indeed deduce the following quantum law for the
shifts of space-time observables
\begin{equation}
\left( \Delta ,X^\mu \right) =-\delta ^\mu (X)-\frac{\partial \widehat{%
\Delta }}{\partial P_\mu }  \label{DX}
\end{equation}
The first classically-looking term is corrected by the second term which
differs from $0$ only for transformations to accelerated frames. In the
particular case of two identical intersecting rays, we obtain this
correction from (\ref{Casimir})
\begin{equation}
-\frac{\partial \widehat{C}_\nu }{\partial P_\mu }=-\frac \alpha {M^2}\left(
\eta _\nu ^\mu -\frac{2P^\mu P_\nu }{M^2}\right)
\end{equation}
Hence, the shifts (\ref{DX}) of space-time observables do not obey the
covariance rules inherited from classical relativity, and the corrections to
these rules follow from conformal algebra. The classical approach presented
in the first part of this letter was an approximation where light rays were
treated as dispersionless distributions. This approximation amounts to set
the Casimir operators to $0$, in which case the correction vanishes so that
observables are transformed as classical parameters. For quantum fields in
contrast, uncertainty relations forbid to set the Casimir operators to $0$
and the transformations of observables effectively differ from those of
coordinate parameters \cite{PRL96}. In particular, space-time observables
are mixed with momentum operators under frame transformations.

For transformations to uniformly accelerated frames, we have found that the
shifts of space-time observables do not have a classically looking form, in
contrast to those of energy-momentum operators. It is however possible to
write down consistency statements for these shifts which allow to make
contact with classical laws. Such relations follow from the fact that the
canonical commutators (\ref{canonical}) are invariant under frame
transformations, since they are classical numbers
\begin{equation}
\left( \Delta ,\left( P^\mu ,X^\nu \right) \right) =0
\end{equation}
It then follows from Jacobi identity that variations of space-time and
energy-momentum shifts are connected through
\begin{equation}
\left( P^\mu ,\left( \Delta ,X^\nu \right) \right) =\left( X^\nu ,\left(
\Delta ,P^\mu \right) \right)
\end{equation}
Both types of variations therefore have a classically-looking form (see (\ref
{DP}) and (\ref{DX}))
\begin{equation}
-\frac \partial {\partial X_\mu }\left( \Delta ,X^\nu \right) =\frac \partial
{\partial P_\nu }\left( \Delta ,P^\mu \right) =\partial ^\mu \delta ^\nu (X)
\label{DDX}
\end{equation}

In the particular case $\mu =\nu =0$, the first quantity is the shift of a
clock rate, while the second quantity is the change of the redshift of
frequency. Relations (\ref{DDX}) express both quantities in terms of the
function $\partial ^0\delta ^0$ that is also the metric coefficient $g^{00}$
in equation (\ref{cfac}). These results prove that the relativistic
transformations of space-time scales and energy-momentum shifts are now
consistently obtained in the quantum domain from the same commutation
relations. The close connection between relations (\ref{DDX}) and the metric
tensor has in fact more general implications. Quantum analogs of the
definition (\ref{cfac}) of the conformal factor are indeed obtained by
symmetrising expression (\ref{DDX}) in the exchange of the two indices $\mu $
and $\nu $
\begin{eqnarray}
\frac \partial {\partial X_\mu }\left( \Delta ,X^\nu \right) +\frac \partial
{\partial X_\nu }\left( \Delta ,X^\mu \right)  &=&2\eta ^{\mu \nu }\lambda
(X)  \nonumber \\
\frac \partial {\partial P_\mu }\left( \Delta ,P^\nu \right) +\frac \partial
{\partial P_\nu }\left( \Delta ,P^\mu \right)  &=&-2\eta ^{\mu \nu }\lambda
(X)  \label{Dl}
\end{eqnarray}
The classical definition (\ref{cfac}) was written in terms of classical
coordinate parameters and it was therefore not properly defined from an
operational point of view. In contrast, the quantum definition (\ref{Dl}) is
expressed in terms of observables and may thus be deduced from measurements
of field quantities. It is a remarkable consequence of the consistency
statement discussed in the previous paragraph that the conformal factor may
be deduced from measurements of space-time or energy-momentum observables.
It is also remarkable that the conformal factor appears in the
transformation laws (\ref{Dl}) of space-time observables, although
propagation of electromagnetic field is known to be insensitive to a
conformal variation of the metric tensor \cite{MG80}.

In the present letter we have introduced quantum observables describing the
space-time position of the physical event defined by the intersection of two
light pulses and we have shown that these observables are canonically
conjugated to energy-momentum operators associated with the same field
state. In particular, a time observable has been defined which is
canonically conjugated to energy. The shifts of these observables under
transformations to accelerated frames have been derived from conformal
algebra and shown to depart from the laws of classical relativity, since
space-time and energy-momentum are mixed. The Einstein laws for
transformations of clock rates and frequency redshift have nevertheless been
recovered and expressed in terms of the conformal factor.

In more general words, these results introduce a new conception of space
built on the notion of quantum points. These points are the analogs in
quantum theory of the localised events of relativity theory. The
relativistic properties of this space are determined by the conformal
symmetry which underlies quantum theory of massless fields. The metric
coefficients which may be deduced from measurements of space-time or
energy-momentum observables must also be considered as quantum operators.
These results clearly point to the need for a quantum geometry \cite{Connes}.

\end{document}